\documentclass[pdflatex,sn-mathphys-num]{sn-jnl}

\usepackage{graphicx}%
\usepackage{multirow}%
\usepackage{amsmath,amssymb,amsfonts}%
\usepackage{amsthm}%
\usepackage{mathrsfs}%
\usepackage[title]{appendix}%
\usepackage{xcolor}%
\usepackage{textcomp}%
\usepackage{manyfoot}%
\usepackage{booktabs}%
\usepackage{algorithm}%
\usepackage{algorithmicx}%
\usepackage{algpseudocode}%
\usepackage{listings}%

\usepackage[export]{adjustbox}

\newcommand{\ket}[1]{|#1\rangle}
\newcommand{\bra}[1]{\langle#1|}
\newcommand{\braket}[2]{\langle#1|#2\rangle}
\DeclareMathOperator{\Tr}{Tr}
\newcommand{\IVZ}{[\textit{Approximate $Z$ condition}]}
\newcommand{\CRF}{[\textit{RG flow condition}]}
\newcommand{\UPF}{[\textit{UV-free condition}]}

\theoremstyle{thmstyleone}%
\theoremstyle{thmstyletwo}%

\theoremstyle{thmstylethree}%

\begin{document}

\title[Article Title]{
    Essential difference between 2D and 3D from the perspective of real-space renormalization group
}


\author*[1,3]{\fnm{Xinliang} \sur{Lyu}}\email{xlyu@ihes.fr}

\author[1,2]{\fnm{Naoki} \sur{Kawashima}}\email{kawashima@issp.u-tokyo.ac.jp}

\affil*[1]{\orgdiv{Institute for Solid State Physics}, \orgname{The University of Tokyo}, \orgaddress{\street{Kashiwanoha 5-1-5}, \city{Kashiwa}, \postcode{277-8581}, \state{Chiba}, \country{Japan}}}

\affil[2]{\orgdiv{Trans-scale Quantum Science Institute}, \orgname{The University of Tokyo}, \orgaddress{\street{Hongo 7-3-1}, \city{Tokyo}, \postcode{113-0033}, \country{Japan}}}

\affil[3]{\orgname{Institut des Hautes \'Etudes Scientifiques (IHES)}, \orgaddress{\street{35, route de Chartres}, \city{Bures-sur-Yvette}, \postcode{91440}, \country{France}}}


\abstract{
We point out that area laws of quantum-information concepts indicate limitations of block transformations as well-behaved real-space renormalization group (RG) maps, which in turn guides the design of better RG schemes.
Mutual-information area laws imply the difficulty of Kadanoff's block-spin method in two dimensions (2D) or higher due to the growth of short-scale correlations among the spins on the boundary of a block.
A leap to the tensor-network RG, in hindsight, follows the guidance of mutual information and is efficient in 2D, thanks to its mixture of quantum and classical perspectives and the saturation of entanglement entropy in 2D.
In three dimensions (3D), however, entanglement grows according to the area law, posing a threat to a 3D block-tensor map as an apt RG transformation.
As numerical evidence, we show that estimates of 3D Ising critical exponents fail to converge with respect to the RG step, making the 3D block-tensor map an unreliable RG method.
Moreover, the estimates do not improve as more coupling constants are retained.
As a guidance to proceed, a tensor-network toy model is proposed to capture the 3D entanglement-entropy area law.
}




\maketitle

\section{Introduction}\label{sec:intro}
The renormalization group (RG) idea is an overarching theoretical framework for both elementary particle physics and statistical physics of complex systems~\cite{wilson-kogut1974, wilson1975kondo}.
It provides a modern view of the regularization process of removing infinities in quantum fields of elementary particles;
and the concepts of RG relevant and irrelevant directions have deepened our understanding of universality and emergent phenomena ubiquitous in complex systems.
The essence of the RG idea is the separation of universal physics from microscopic details using a divide-and-conquer strategy.
A judicious cleanse of microscopic details is the key for designing systematically-improvable RG techniques because the computational power can be saved for universal physics.
Wilson's momentum-shell RG combined with $\epsilon$-expansion~\cite{wilson-fisher1972} is the first demonstration of the power of this separation of scales, organized according to the Fourier modes of Landau-Ginzburg-Wilson field.
However, for non-perturbative approaches, like real-space RG, the design of a good RG transformation is more of an art than a science~\cite{fisher1998}.

\par
In this paper, we want to turn some aspects of the design of real-space RG from an art into a science by articulating the following questions:
What properties should an apt RG transformation have?
Is there a separation of system-scale and lattice-scale physics in real-space RG\@?
If so, where are they?
Is it necessary for a real-space RG transformation to integrate out all lattice-scale physics, or is it just optional?
We demonstrate in the following that the area laws of quantum-information concepts have a lot to say about these questions.

\par
How do area laws have anything to do with real-space RG transformations?
The connection is due to the fact that a block transformation, as the starting point of many real-space RG, generates increasingly larger blocks of the original system, which encapsulates ``information'' scaling according to area laws.
An example of how the entanglement-entropy area law can guide the design of an efficient RG map for a quantum chain is Entanglement Renormalization invented by Vidal~\cite{vidal2003,Vidal:2007,Vidal:2010:ER}.
Entanglement in a ground state $\ket{0}$ can be quantified in the following way:
divide the system into two parts: a block $B$ and its complement $\bar{B}$; define the density matrix of $B$ by tracing out the Hilbert space of $\bar{B}$ as
\begin{align}
    \label{eq:rho-def}
    \rho_B = \Tr_{\bar{B}}\left(\ket{0}\bra{0}\right);
\end{align}
and the entanglement between the block $B$ and the remaining system $\bar{B}$ can be measured by von Neumann entropy of $B$: 
\begin{align}
    \label{eq:ee-def}
    S(\rho_B) = -\Tr_B\left(\rho_B \log \rho_B\right).
\end{align}
An area law means a scaling behavior like the boundary area $|\partial B|$ of the block: $S(\rho_B) = O(|\partial B|)$ in the case of entanglement entropy~\cite{Eisert2010area}.
In a (1+1)D gapped system, the boundaries are two points and entanglement entropy becomes a constant for a block $B$ with a size larger than the correlation length;
this saturation of entanglement entropy justifies the DMRG approximation of keeping only a constant number of states of the block~\cite{vidal2003}.

    We emphasize here that the two parts of the system are divided in \emph{real space}.
    The amount of entanglement depends drastically on the space where the state is represented.
    For example, a non-interacting fermionic system can be completely factorized in momentum space, but obeys an area law in real space, sometimes with a logarithmic correction~\cite{Wolf:2006}.
    One can try circumventing the area law in real space by analyzing models expressed in momentum space, like lowest Landau level orbits on a two-sphere~\cite{FuzzySphere:2023}.

\par
The focus here is on the properties of real-space RG in the context of classical statistical mechanics, specifically about block transformations, using area laws as a tool.
One hopes that a better understanding of the nature and limitations of such a basic operation can guide the design of efficient and well-behaved real-space RG transformations.
    We restrict our discussion to these real-space RG transformations where the RG map is constructed directly by imposing the condition that the partition function should be well approximated.
    Examples of such kind include Migdal-Kadanoff (MK) bond-moving approach~\cite{Migdal:1975a,Kadanoff:1976}, Wilson's numerical implementation of a block-spin map~\cite{wilson1975kondo} and other variations of MK scheme~\cite{Kadanoff:1975:var,Martinelli:Parisi:1981}.
    The modern tensor-network renormalization group (TNRG) methods~\cite{levin-nave} also belong to this type; they can be seen as a generalization of the MK approach~\cite{Meurice:2013}. 
    In this sense, the Monte Carlo RG~\cite{Swendsen:1979} does not fit in the scope of this paper.
    The reason is that the Monte Carlo RG involves a stochastic sampling process; it complicates the discussion here, where the emphasis is the properties of an RG map per se.

    The argument in this paper is inspired by Levin and Nave~\cite{levin-nave}. 
    When implementing the first block transformation in tensor-network language, they use an area-law argument to show that the numerical truncations in their scheme cannot be guaranteed to be accurate in 2D critical and 3D systems.
    They see the TNRG as a numerical method for approximating the partition function.
    We expand their argument to show further implications of the area law for understanding the properties of TNRG in the context of the Wilsonian RG framework.
    Our argument helps clarify why the difficulty of the TNRG in 3D is qualitatively different from 2D, as well as some confusions in the 3D TNRG calculations~\cite{hotrg2012,wang2014,jha-3dpotts}.

\section{Properties of an apt RG}\label{sec:RGapt}
Before assessing a real-space RG, it is better to have a clear-cut answer to the first question:
What is a good RG transformation in a statistical physics setting?
We list three basic properties: 

    {\IVZ}: The partition function $Z$ after an RG map should be close to the original one. 

    {\CRF}: An RG map should exhibit fixed points and reproduce well-known RG flows.

    {\UPF}: An RG map should strive to identify and filter out short-scale physics as much as possible, so the finite computation power and memory are saved for the universal properties.

\par
The first {\IVZ} is the defining property of a RG transformation in statistical physics.
Suppose we have the reduced energy of a system as
$\mathcal{H}_{\{K_j\}}[\{s_{\mathbf{r}}\}]$, where $s_{\mathbf{r}}$
denotes a spin located at lattice site
$\mathbf{r}$ and $K_j$ is the $j$-th coupling constant parametrizing the energy function $\mathcal{H}$. 
The statistical weight of a given configuration of spins is the Boltzmann factor
$w(\{K_j\}; \{s_{\mathbf{r}}\}) = e^{-\mathcal{H}_{\{K_j\}}[\{s_{\mathbf{r}}\}]}$.
The partition function $Z\left(\{K_j\}\right) = \sum_{\{s_{\mathbf{r}}\}} w(\{K_j\}; \{s_{\mathbf{r}}\})$,
as a function of coupling constants, generates various correlation functions.
A real-space RG transformation is a rule of mapping the original spins $\{s_{\mathbf{r}}\}$ to renormalized ones $\{s'_{\mathbf{r}'}\}$,
inducing a map in coupling-constant space $\mathcal{T}: \{K_j\} \mapsto \{K_j'\}$, such that the new system with energy function
$\mathcal{H}_{\{K_j'\}}[\{s'_{\mathbf{r}'}\}]$ 
has the same partition function as the old one
$Z\left(\{K_j'\}\right) = Z\left(\{K_j\}\right)$.
In a practical calculation, the partition function in the new description should be a good approximation of the original one.
A direct consequence of {\IVZ} is that all correlation functions between renormalized spins in larger scales are encoded in $Z\left(\{K_j'\}\right)$ of a coarser system without knowing anything about the original one~\cite{cardy-book}.

\par
The second property, {\CRF}, makes it possible to identify different phases of matter from different isolated fixed points under the RG map $\mathcal{T}:\{K_j^*\} \mapsto \{K_j^*\}$;
universal properties like critical exponents can then be extracted from the linearized RG transformation at a critical fixed point.
One may wonder whether {\CRF} is related to the defining property {\IVZ}.
In practice, truncation is usually necessary.
It is possible that an unwise truncation scheme causes incorrect RG flows.
However, we want to point out that truncation is \textit{not the only} cause of incorrect RG flows.
    In fact, Kadanoff~\cite{Kadanoff:1976} argued that the spin decimation, even if performed exactly, leads to a spin-spin correlation that does not change under a change of length scale;
    this indicates that the spin decimation always produce fixed points with the exponent $\beta / \nu = \Delta_{\sigma} = 0$.
Recently, it is demonstrated that an exact block transformation in tensor-network language also gives unsatisfactory RG flows~\cite{kennedy-hiT, kennedy-lowT}, where both the high- and the low-temperature fixed points form a continuum, instead of isolated fixed points.

\par
The above two properties are the bare minimum for an apt RG transformation at a conceptual level.
In a practical calculation, only a finite number of couplings are retained.
An effective RG retains couplings related to universal long-wavelength physics.
The last property {\UPF} is for this task.
Moreover, we will see in the following area-law argument that this property is closely related to the first two.
Now we are ready to assess various real-space RG schemes.

\section{Kadanoff's block spin}\label{sec:blockspin}
We go back all the way to the origin of real-space RG and examine Kadanoff's block-spin idea~\cite{kadanoff-spinblock}.
A block-spin method, like decimation, if performed exactly, entails an indefinite growth of couplings.
Can the usual practice of retaining a constant number of couplings be justified?
%
Mutual information and its area law for a system in thermal equilibrium shed some light on this question~\cite{verstraete2008, verstraete-talk}.
Intuitively, the correlation between the block of spins $B$ and its environment $E$ is mediated through a strip of spins around their interface, roughly with the width of a characteristic length scale $\xi$ of the system~\cite{verstraete2008}.
This correlation can be quantified using mutual information $I(B, E)$ and is bounded by the number of spins inside the strip, $I(B, E) \leq 4 |\partial B| \xi$, 
where $|\partial B|$ is the area of the boundary of the block.
In the worst case, for a classical statistical system in $d \geq 2$, the correlations between the block and its environment increase according to $L^{d-1}$ as the size $L$ of the block grows under RG\@. 
The tension between the \textit{growth} of mutual information and the \textit{constant} number of couplings retained during RG suggests one possible explanation of why many real-space RG schemes based on spin pictures fail to produce systematically improvable estimates of critical exponents~\cite{kardar-book,shankar-qftbook}.

\par
Compared with the indefinite growth of couplings, the mutual-information area law reveals something more interesting.
The upper bound $|\partial B | \xi$ is due to microscopic physics arising from spins in the strip, resulting in a risk of violating {\UPF} in block-spin methods.
The argument thus suggests a direction for a better real-space RG design,
a leap from real-space RG in spin pictures to that in tensor-network representations.

\section{Block tensor and entanglement entropy}\label{sec:bktenEE}
The two essential ideas of tensor-network renormalization group (TNRG) are~\cite{levin-nave}

(1) focusing on spins at the boundary of a block and

(2) a mixture of quantum and classical view, so we can use unitary transformations to identify important states.

\par
For a large block, the partition function with boundary spin fixed while inner spins summed over represents the ground state $\ket{g}$ of a quantum Hamiltonian that corresponds to the transfer matrix acting on the boundary spins~\cite{shankar-qftbook,levin-nave,kogut1979}.
The advantage of the quantum perspective is that area laws of entanglement for quantum systems become applicable to classical statistical systems, with two reductions of dimensionality---one from classical-to-quantum mapping and the other from an area law.
%
One salient feature of the TNRG is that scaling dimensions of a variety of scaling operators are contained in the linearized tensor RG map, if the linearization is performed in a particular manner~\cite{lyu-kawashima2021,Ebel:2024LDO}.
For example, in Fig.~\ref{fig:est-scaleD}, the linearized RG map contains scaling dimensions of the energy-momentum tensor $T_{ij}$ and various descendants like $\partial_i \epsilon, \partial_i \sigma$ with $i = x,y,z$ of the 3D Ising criticality, most of them with an approximate degeneracy anticipated by conformal field theory.
This has not been demonstrated in previous RG methods.

\par
Specifically, imagine a 2D classical system with partition function $Z$ and divide it into square blocks.
Each block, as a $4$-leg tensor, represents the ground state wavefunction~\cite{levin-nave} of spins located at the boundary,
\begin{align}
    \label{eq:block2d}
\braket{\vec{s}_l, \vec{s}_r, \vec{s}_u, \vec{s}_d}{g} = A_{\vec{s}_l,\vec{s}_r, \vec{s}_u, \vec{s}_d} 
= 
\includegraphics[scale=0.8, valign=c]{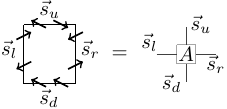}\quad,
\end{align}
where, for example, $\vec{s}_{l}$ denotes all spins on the left edge of the square. 
With copies of the block as tiles, the partition function $Z$ can be represented by the contraction $Z_N(A)$ of a tensor network of $N$ copies of $A$, so $Z = Z_N(A)$~\cite{levin-nave}.
%
    The density matrix of a leg of the tensor $A$ is obtained according to Eq.~\eqref{eq:rho-def} by choosing a leg as the system $B$ while the remaining legs as $\bar{B}$.
    For example, if one chooses leg $\vec{s}_l$, the density matrix is\footnote{
    Notice that the complex conjugate is not needed for classical statistical models since the tensor is real-valued.
    }
    \begin{align}
        \label{eq:denMatA}
        \rho_{\vec{s}_l' \vec{s}_l} =
        \sum_{\vec{s}_r, \vec{s}_u, \vec{s}_d} 
        A_{\vec{s}_l',\vec{s}_r, \vec{s}_u, \vec{s}_d}
        A_{\vec{s}_l,\vec{s}_r, \vec{s}_u, \vec{s}_d},
    \end{align}
    from which the entanglement entropy can be calculated according to Eq.~\eqref{eq:ee-def}.
    \emph{
    This single-tensor-level entanglement entropy serves as a single-number characterization of the tensor, which can help understand {\CRF} of a block-tensor map.
    }
    Similarly, a density matrix can be obtained for a larger patch of the tensor network.

\begin{figure}[tb]
    \centering
    \includegraphics[scale=1.0,
    valign=c]{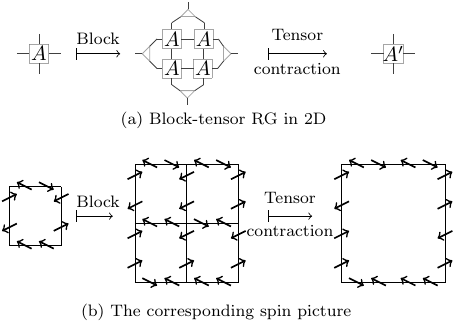}
    \caption{\label{fig:bkten}
        Block-tensor transformation and its spin picture with $b=2$.
        The ``Block'' step in this Figure is the Step (i), while the ``Tensor contraction'' step corresponds to Step (ii) and (iii).
        The triangular tensor in (a) is the unitary transformation that identifies important states.
        In numerical calculations, truncations of the state are according to the eigenvalue spectrum of the density matrix of the larger square block after the ``Block'' step.
        When truncations happen, the triangular tensors are known as isometric tensors~\cite{tnr2015}.
}
\end{figure}

The 2D block-tensor transformation with rescaling factor $b$ has three steps (Fig.~\ref{fig:bkten} shows a $b=2$ transformation)---

Step (i): grouping a larger square block of $b^2$ copies of tensor $A$, 

Step (ii): summing over spins on inner edges, and 

Step (iii): a unitary transformation on spins sitting on each outer edge of the larger square block 
\textit{to identify important states according to the eigenvalue spectrum of the density matrix of the larger square block} (see Fig.~\ref{fig:bkten}).

These three steps are general in higher dimensions.
In 2D, the Hilbert space corresponding to the larger square block is similarly divided into $4$ groups according to $4$ edges of the larger square block, which defines a coarse tensor $A'$
This defines a map in tensor space,
 \begin{align}
    \label{eq:tnrg-map}
    \mathcal{T}: A \mapsto A'.
\end{align}
We have $Z_N(A) = Z_{N/b^2}(A')$ by construction, provided no truncation made in Step (iii). 
%
    From Fig.~\ref{fig:bkten}, we see that under a block-tensor transformation, the size of the spin block that the coarse-grained tensor $A'$ represents becomes $b$ times as large as what $A$ represents. 

    In numerical calculations, usually the first $\chi$ eigenstates of the density matrix of the larger square block with the $\chi$ largest eigenvalues are chosen, the remaining states being truncated.
    The error of this truncation can be approximated by the $\chi$-th eigenvalue if the density matrix is normalized to have unit trace.
    Recall that a larger von Neumann entropy of an eigenvalue spectrum usually corresponds to a slow decay of the spectrum, thus indicating a larger RG truncation error.
    \emph{
    Therefore, the second role of entanglement entropy is how well the partition function is preserved under the RG transformation, reflecting {\IVZ} of a block-tensor map.
    }

In summary, the role of entanglement entropy in the block-tensor map is twofold---it implies how well the partition function is preserved and also serves as a single-number characterization of the tensor, reflecting {\IVZ} and {\CRF} respectively.
A larger entanglement entropy indicates a larger RG truncation error, thus a poor RG approximation for the partition function.
A changing of the entanglement entropy with respect to the size of the system, or the RG step, indicates the tensor itself is not fixed under the RG transformation.
These two roles of entanglement entropy will be employed to understand the properties of the block-tensor map in 2D and 3D in the following section.

\section{Is the block-tensor map an apt RG?}\label{sec:bktenRG}
\subsection{In 2D}
According to the entanglement area law of (1+1)D gapped systems, the density matrix $\rho_{\vec{s}_l}$ of the subsystem $\vec{s}_l$ has entanglement entropy $S(\rho_{\vec{s}_l}) \to S_0 \sim \log\xi$ if the statistical system has characteristic length scale $\xi$ and the block size $L \gg \xi$~\cite{vidal2003,cardy-eeqft-intro}.
As a result, as long as the number of states $\chi$ retained in Step (iii) is $O(\xi)$, the approximation is quasi-exact~\cite{levin-nave}.

\par
The above arguments imply that the 2D block tensor can be an apt RG transformation.
{\IVZ} is well satisfied due to the saturation of entanglement entropy. 
This is a great step forward from Kadanoff's block-spin method.
This advance is achieved thanks to the aid of density matrix $\rho_{\vec{s}_l}$ for the identification of short-scale correlations among spins $\vec{s}_l$ on an edge of a square block, a manifest improvement on {\UPF}.
However, $\rho_{\vec{s}_l}$ still preserves some short-scale correlations between the spin group $\vec{s}_l$ and the rest $\vec{s}_u \cup \vec{s}_d \cup \vec{s}_r$, which is reflected in $S_0 \sim \xi$.
The remnant short-scale entanglement is responsible for continua of fixed-point tensors in the neighborhood of trivial phases~\cite{levin-nave,gu-wen2009}, which was noticed since the birth of 2D TNRG methods---{\CRF} is violated.
Efforts to filter the remnant entanglement lead to beyond-simple-block 2D TNRG schemes with a \textit{full satisfaction of all three properties}, as numerical methods~\cite{gu-wen2009,tnr2015,tnr-plus2017,loop-tnr2017,gilt2018,harada2018,fet2018} and as exact RG transformations~\cite{kennedy-hiT,kennedy-lowT}.
%

\subsection{Exact 3D block-tensor map}

The advantage of TNRG in 2D makes it a promising candidate as a 3D real-space RG\@.
One thing to note is that, without the viewpoint of entanglement entropy and its area laws, whether the block-tensor RG in 3D is better or worse than that in 2D is not straightforward.
Recall that in Wilson's momentum-shell RG, systems in dimensions $d > 4$ are easy, where mean-field treatment of Landau-Ginzburg field is enough; but in $d < 4$, perturbative $\epsilon$-expansion is necessary and $d = 2$ is harder than $d = 3$ since $\epsilon = 4 - d$ is larger.
Interestingly, numerical experiments~\cite{wang2014,jha-3dpotts} would seemingly imply that block-tensor RG in 3D improves {\CRF} compared to 2D.
However, the truncation effects, as a cause of those observations, become more transparent through the lens of entanglement entropy.

\begin{figure}[tb]
    \centering
    \includegraphics[width=0.6\columnwidth,
    valign=c]{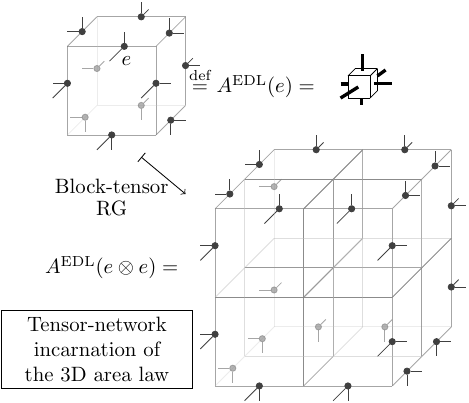}
    \caption{\label{fig:edl}
    (Upper left)
    A tensor $A^{\text{EDL}}(e)$ with an EDL structure is the tensor product of 12 copies of edge-matrix $e$ located at 12 edges of a cube block; 
    every 4 matrix-indices on the same face are grouped together to be a thicker leg of the 6-leg $A^{\text{EDL}}(e)$ .
    (Lower right)
    A block-tensor RG with rescaling factor $b=2$.
A total of $12 \times 8 = 96$ copies of edge-matrix $e$ form 3 types:
(i) Around each of the 6 axes inside the block, 4 copies of $e$ form a loop and becomes a number, so there are $4 \times 6$ copies;
(ii) On each of 6 faces, copies of $e$ form 4 entanglement pairs inside the face, so there are $2\times 4 \times 6$ copies;
(iii) The remaining $2\times 12$ copies sit on the 12 edges of the larger block, each edge having 2 copies side by side.
Only type-(iii) copies survive the exact block-tensor RG; the other two types become an overall multiplication factor. 
The coarser tensor has an EDL structure built from $e \otimes e$: $A' = A^{\text{EDL}}(e \otimes e)$. 
}
\end{figure}

\par
The assessment of the block-tensor RG in 3D parallels its analysis in 2D.
We have a 6-leg tensor $A_{\vec{s}_{x+}, \vec{s}_{x-}, \vec{s}_{y+}, \vec{s}_{y-}, \vec{s}_{z+}, \vec{s}_{z-}}$ (for example, $x+$ denotes the leg along the positive $x$ axis), corresponding to a cubic block of spins.
The TNRG map in Eq.~\eqref{eq:tnrg-map} satisfies $Z_N(A) = Z_{N/b^3}(A')$.
For a large block with linear size $L$, the area law of 3D quantum field theories says that the entanglement entropy of a subsystem on a face of the cube, $S(\rho_{\vec{s}_{x+}})$ for example, scales like~\cite{kitaev-preskill,nishioka2018} 
\begin{align}
    \label{eq:area3dcft}
    S(\rho_{\vec{s}_{x+}})
    =
    \alpha L - F,
\end{align}
where $\alpha$ contains microscopic details and $F$ encodes universal properties of the system.
This entropy should not be confused with the mutual information between the inside and the outside of the cube mediated through the face.
Rather, it is from the correlations between the two surface areas mediated through the edge region.
%
The area law in Eq.~\eqref{eq:area3dcft} reveals a lot about 3D block-tensor transformation.
The \emph{growth} of $S(\rho_{\vec{s}_{x+}})$ with the size $L$ of the block marks a \emph{qualitative difference} between 3D and 2D block tensor RG\@.
Retaining a constant number of couplings in 3D loses its justification even off the criticality, which works in 2D due to the saturation of $S \to S_0$;
this poses a threat to {\IVZ}. 
More details come later.
Even if the block-tensor RG is exact and {\IVZ} is strictly satisfied, the scaling behavior of entanglement bodes ill for {\CRF}.
Let's say a block-tensor scheme is applied to the nearest-neighbor Ising model on a cubic lattice, how does the tensor RG flow look like?
At infinite temperature, or $\beta = 1/T = 0$, after a gauge transformation~\cite{kennedy-lowT}, the tensor $A \to A^{\text{hi}}$ only has a single nonzero component, $A^{\text{hi}}_{000000} = 1$\footnote{
        For this high-$T$ fixed-point tensor,  the index value 0 means an equal-weight superposition of all the spin-based states represented by this leg.
        For example, if a leg of the initial tensor corresponds to a single spin, then the state 0 represents $(\ket{\uparrow} + \ket{\downarrow}) / \sqrt{2}$.
        This also applies to the high-$T$ edge matrix $e^{\text{hi}}$ in the discussion below Eq.~\eqref{eq:rgEDL}.
};
the entanglement entropy $S(\rho_{\vec{s}_{x+}})$ vanishes so $\alpha = F = 0$ in Eq.~\eqref{eq:area3dcft}.
Now we consider $0 < \beta \ll 1$.
 The entanglement $S(\rho_{\vec{s}_{x+}})$ of the resultant tensor has a small but non-vanishing $\alpha \neq 0$ in Eq.~\eqref{eq:area3dcft}.
A well-behaved RG transformation should generate an RG flow towards the high-temperature fixed point $A^{\text{hi}}$ with zero entanglement entropy, because the Ising model with $\beta \ll 1$ is in high-temperature phase.
However, Eq.~\eqref{eq:area3dcft} shows that the coarser tensors generated by the block-tensor scheme has growing entanglement entropy due to the growth of $L$ and an $\alpha \neq 0$.
Therefore, the 3D block-tensor RG, even if performed exactly by retaining infinite states, fails to satisfy {\CRF} in the vicinity of the high-temperature fixed point.
Notice that the failure is not due to any approximation.
Instead, as is indicated by the area law in Eq.~\eqref{eq:area3dcft}, the incorrect RG flow is closely related to the violation of {\UPF}---the diverging term $\alpha L$ arises from microscopic details reflected in $\alpha$.
%
At criticality, the growth of the entanglement entropy suggests that an exact 3D block-tensor RG cannot exhibit a critical fixed-point tensor, since the entanglement entropy is a single-number characterization of the tensor.
This bodes ill for the hope of obtaining a critical fixed-point tensor numerically using a 3D block-tensor map\footnote{
    See Ref.~\cite{Lyu:Kawashima:2024} for numerical evidence.
}.

\par
The area law in Eq.~\eqref{eq:area3dcft} has an incarnation as a tensor-network toy model that mimics the entanglement structure of the high-temperature phase.
The toy model captures the remnant entanglement as matrices sitting on the boundary.
It is known in 2D as a corner-double line (CDL) structure~\cite{gu-wen2009}.
We call its 3D generalization an edge-double-line (EDL) structure, since the boundaries between faces of a cube are edges, see Fig.~\ref{fig:edl};
the EDL structure transforms under the exact block-tensor RG with rescaling factor $b$ as,
\begin{align}
    \label{eq:rgEDL}
    A^{\text{EDL}}(e) \mapsto A' = A^{\text{EDL}}(e^{\otimes b}).
\end{align}
The coarser $A'$ has 1-face entanglement entropy $S(\rho^{\text{EDL}}_{\vec{s}_{x+}}) = 4S_e b$,
which grows linearly with block size $b$ with the entanglement entropy $S_e$ of a single edge-matrix $e$ as the pre-factor,
corresponding to $L$ and $\alpha$ in the area law in Eq.~\eqref{eq:area3dcft} respectively.
The correspondence between $\alpha$ and $S_e$ suggests that $e$ captures microscopic-scale physics.
The block-tensor RG leaves copies of $e$ growing under RG so violates {\UPF}.
The high-temperature fixed point $A^{\text{hi}}$ corresponds to $A^{\text{EDL}}(e^{\text{hi}})$, with all $e^{\text{hi}}_{ij}$ vanish except $e^{\text{hi}}_{00} = 1$.
The block-tensor RG fails to simplify $e$ so violates {\CRF}.

\subsection{Numerical 3D block-tensor map}

\par
So far we have focused on the \emph{exact} block-tensor RG in 3D and {\IVZ} is strictly satisfied.
Next, we study the implication of the area law for a practical implementation of 3D block tensor, where we insist on keeping only $\chi$ states during RG.
The failure of {\UPF} reflected in Eqs.~\eqref{eq:area3dcft} and~\eqref{eq:rgEDL} poses a challenge to 3D block-tensor RG as numerical techniques.
Large amounts of computation power and memory are wasted on lattice-scale information $\alpha L$, while the finite and interesting universal information (for example, the 3D Ising universality class has
$F = 0.0623$~\cite{f3dising}) is buried among them.
%

\begin{figure}[tb]
    \centering
    \includegraphics[width=0.8\textwidth,
    valign=c]{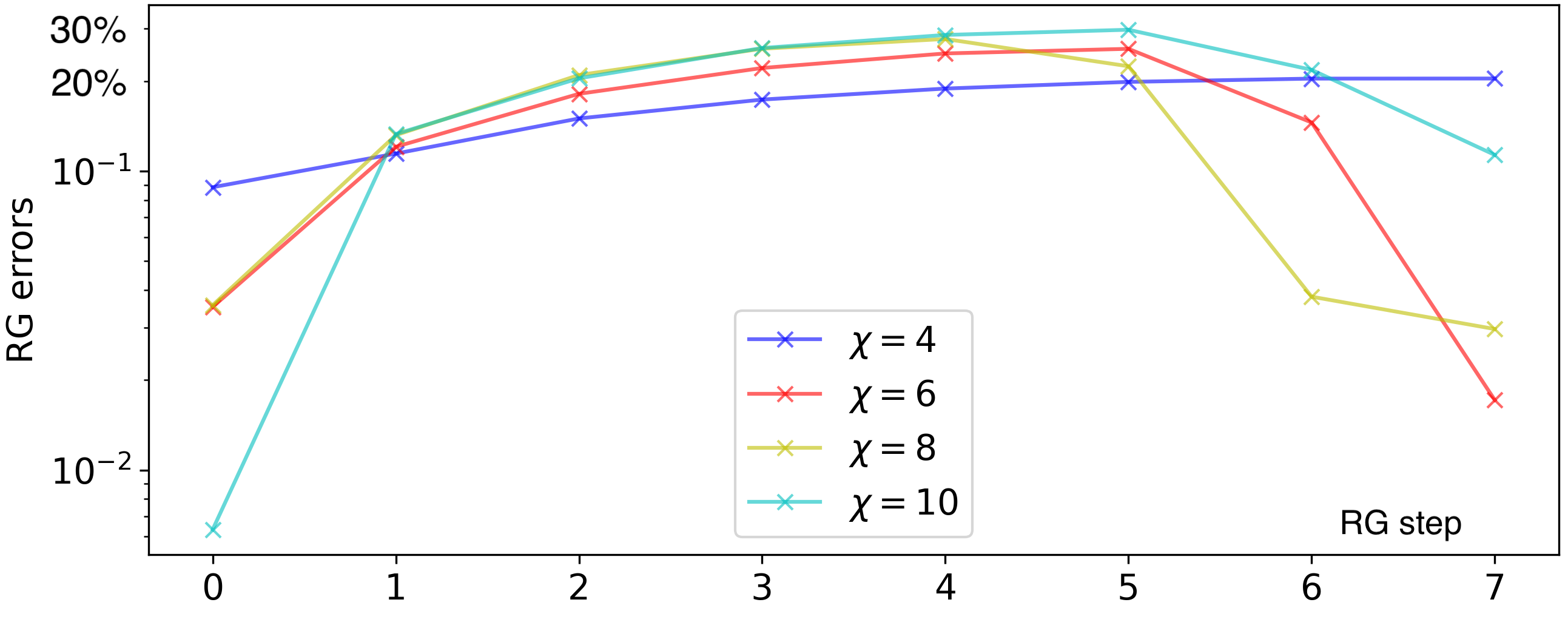}
    \caption{\label{fig:hotrgerr}
            The rapid growth of the RG error with respect to the RG step. 
            The temperature is set to be the estimated critical temperature obtained by studying the RG flows, as is explained in Refs.~\cite{lyu-kawashima2021,Lyu:Kawashima:2024}.
            The RG error grows to more than $10\%$ just after the first RG step.
            Near the critical fixed point, which is around RG steps $3 \leq n \leq 5$, the RG error \emph{increases} from $10\%$ to $30\%$ when the bond dimension increases from $4$ to $10$.
            At the largest bond dimension we reach, $\chi=22$ (not shown in the Figure), the RG truncation error near the critical fixed point has grown to about $38\%$.
}
\end{figure}

The linear growth of the entanglement entropy in 3D indicates a rapid growth of RG truncation errors using a block-tensor map, since the size of the block grows exponentially with the RG step $n$.
This expectation can be checked by plotting RG truncation errors with respect to the RG step.
We apply the higher-order tensor renormalization group (HOTRG)~\cite{hotrg2012} to the 3D Ising model with the nearest-neighbor interaction at the estimated critical temperature.
A single RG step in the HOTRG consists of three collapses in $x,y$ and $z$ directions, each collapse having two truncation errors.
In total, there are $2 \times 3 = 6$ truncation errors for a single RG step.
We use the maximum among these six errors to represent the truncation error of a single RG step, and plot how it changes with the RG step $n$ at bond dimensions $\chi \leq 10$ in Fig.~\ref{fig:hotrgerr}.
The block-tensor map in 3D loses its justification as a reliable real-space RG transformation due to this rapid growth of the RG truncation errors\footnote{
        The HOTRG~\cite{hotrg2012} may sometimes seem to produce estimates of scaling dimensions with relatively high accuracy for a ``magic'' bond dimension $\chi=14$.
        We want to point out that the seemingly high accuracy could be an accident that is not robust against the change of the bond dimension.
        This accidental high accuracy is also observed in our calculations in Fig.~\ref{fig:est-scaleD}, where the relative error of the energy-density scaling dimension $\Delta_{\epsilon}$ is smaller than $10^{-2}$ at $\chi=12$.
    }.


\begin{figure}[tb]
    \centering
    \includegraphics[width=1.0\textwidth,
    valign=c]{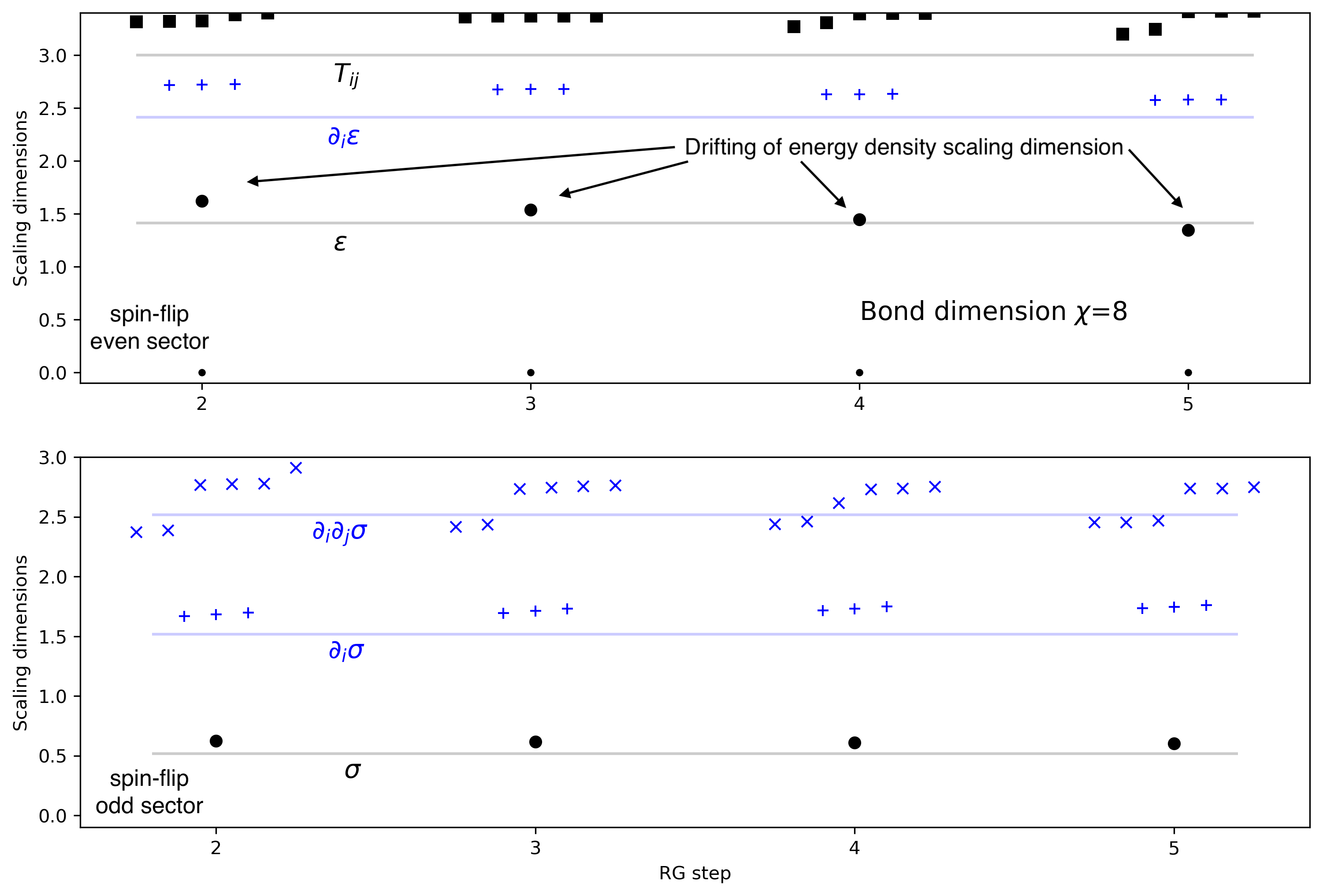}
    \caption{\label{fig:x8scD}
            Drifting of the estimates of scaling dimensions that happens at most bond dimensions $\chi \leq 22$.
            The example shown here is at $\chi = 8$.
            The spin and energy density operators are denoted as $\sigma, \epsilon$, respectively, while $\partial_i \sigma, \partial_{i} \partial_j \sigma$ and $\partial_i \epsilon$ are their descendant operators.
            The energy-momentum operator is denoted as $T_{ij}$.
            The estimate of $\epsilon$ drifts with respect to the RG step.
}
\end{figure}

\begin{figure}[tb]
    \centering
    \includegraphics[width=1.0\textwidth,
    valign=c]{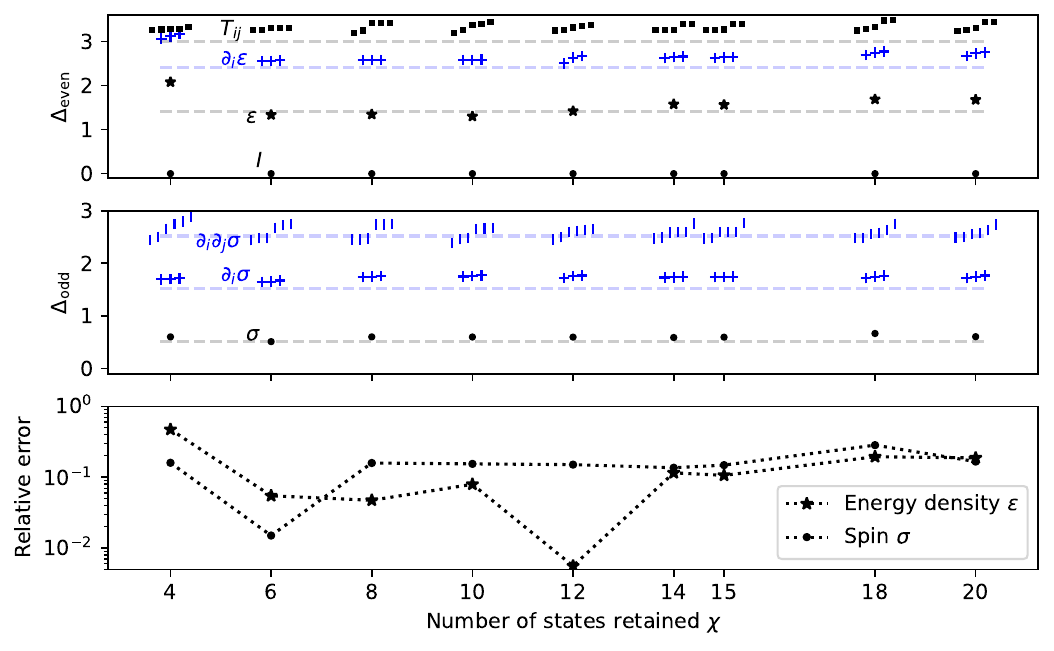}
    \caption{\label{fig:est-scaleD}
        Estimation of 3D Ising scaling dimensions $\Delta$ using high-order tensor renormalization group~\cite{hotrg2012,hotrg-paral2022} and its linearization~\cite{lyu-kawashima2021}.
        Results are organized as per spin-flip $\mathbb{Z}_2$ symmetry.
        Dashed lines are results from conformal bootstrap~\cite{bootstrap-3dising}.
    Most approximation errors of $\sigma$ and $\epsilon$ operators lie between $10\%$ and $20\%$ and fail to decrease with more couplings retained.
}
\end{figure}

    Furthermore, as a single-number characterization of the tensor, the linear growth of the entanglement entropy suggests that it would be unreasonable to expect the block-tensor RG transformation to exhibit a critical fixed-point tensor.
    This poses difficulty in carrying out the Wilsonian RG prescription in a 3D block-tensor RG, where a fixed point is first identified and then the scaling dimensions can be estimated from the linearized RG map at the fixed point (see a short review of the linearized RG method for the HOTRG in Appendix~\ref{secA1:linRGmap}).
The growth of the RG error shown in Fig.~\ref{fig:hotrgerr} demonstrates that the tensor is not fixed near the supposed critical ``fixed point''\footnote{
        More numerical evidence of the fact that the ``fixed-point'' tensor is not fixed can be found in Ref.~\cite{Lyu:Kawashima:2024}.
}.
The consequence is that the estimates of the scaling dimensions using the linearized RG map fail to converge with respect to the RG step (see Fig.~\ref{fig:x8scD}).
Apart from the large RG error near the critical ``fixed point'' shown in Fig.~\ref{fig:hotrgerr}, the drifting of the estimates of the scaling dimensions in Fig.~\ref{fig:x8scD} is another reason that a 3D block-tensor map (like the HOTRG) becomes an unreliable real-space RG method.

    The increase of the RG error with the bond dimension in Fig.~\ref{fig:hotrgerr} suggests that increasing $\chi$ might not be a systematic way to approach the exact solution in practical numerical calculations.
    However, the drifting of the estimates of scaling dimensions in Fig.~\ref{fig:x8scD} makes the study of the dependency of the estimates on $\chi$ trickier.
    One way to decide which RG step to choose for linearizing the RG map is by examining the difference of adjacent tensors and choosing the step in which the tensor is the most stable. 
    Using this strategy\footnote{
        We also tried fixing the RG step for studying the $\chi$ dependency; just like Fig.~\ref{fig:est-scaleD}, we failed to see any systematic trend of the estimates of the scaling dimensions when $\chi$ increases.
    }, we plot the $\chi$ dependency of the scaling dimensions in Fig.~\ref{fig:est-scaleD}.
    This failure of a systematic improvement in a 3D block-tensor map when the bond dimension $\chi$ increases is in a striking contrast to the 2D beyond-simple-block TNRG results\footnote{
    See Fig.~3 in Ref.~\cite{tnr2015} and Table I in the supplementary material of Ref.~\cite{evenbly-vidal-local}.
}.
The numerical results in Figs.~\ref{fig:hotrgerr} to~\ref{fig:est-scaleD} can be reproduced using the {\tt python} codes published in Ref.~\cite{Lyu:bkten3d:codes}.


\par
Curiously, the failure of {\CRF} is hidden in numerical calculations due to a truncation effect.
We can understand this truncation effect by feeding $A^{\text{EDL}}(e)$ into a 3D block-tensor RG with a fixed $\chi$.
Assume without loss of generality that the edge-matrix is positive semi-definite and diagonal $e_{ij} = e_i \delta_{ij}, e_i \geq 0$;
the largest diagonal element is normalized to $e_0 = 1$, and the next largest one is $e_1 < 1$. 
We first perform the exact block-tensor RG with rescaling factor $b$, see Fig.~\ref{fig:edl} and Eq.~\eqref{eq:rgEDL}.
The truncation is based on the eigenvalue spectrum of a 1-face density matrix constructed from $A'= A^{\text{EDL}}(e^{\otimes b})$ (recall that $A'$ can be interpreted as a ground state wavefunction) according to Eq.~\eqref{eq:rho-def}.
The largest eigenvalue is $e_0^{4b} = 1$; the next largest one is $e_1$, with a degeneracy $4b$.
As long as $\chi < 4b$, following the often-adopted numerical practice of throwing away all states in the degenerate subspace, all states except the one with the largest eigenvalue are truncated;
the edge-matrix $e$ is renormalized to $e^{\text{hi}}$.
Therefore, under the approximated block-tensor RG, $A^{\text{EDL}}(e) \mapsto A' = A^{\text{hi}}$ due to this truncation effect.
%
The ``seemingly correct RG flow'' is observed in previous numerical practices~\cite{wang2014,jha-3dpotts}, as well as in the numerical calculation performed in this paper.
%
The caveat is that keeping a constant number of states leads to a bad approximation, which potentially undermines the validity of the 3D block-tensor RG\@.
The seemingly satisfaction of {\CRF} is achieved at the cost of violating {\IVZ}.

\section{Conclusion and remarks}\label{sec:conclusion}
We have demonstrated that the entanglement-entropy area laws offer a unified framework for understanding the properties of the block-tensor map in TNRG.
They indicate how large the RG approximation errors are and how they change with respect to the RG step.
They help diagnose whether the block-tensor map leads to correct RG flows, or specifically, whether it exhibits any critical fixed-point tensor.
Moreover, it indicates the location of microscopic information, and thus can help guide the design of a more efficient TNRG map.

In 3D, the entanglement-entropy area law is an impediment to a block transformation in tensor-network representation.
A similar difficulty of a logarithmic growth of entanglement~\cite{vidal2003} arises in 2D criticality.
Many advanced 2D TNRG schemes~\cite{gu-wen2009,tnr2015,tnr-plus2017,loop-tnr2017,gilt2018,harada2018,fet2018} applicable to criticality have been proposed, all striving to filter the remnant entanglement.
Yet an entanglement-filtering-enhanced 3D TNRG scheme is still missing,
even though the linear growth of entanglement in 3D systems, no matter critical or not, is worse.
%
After the proposal of the first 3D block-tensor scheme, the higher-order tensor renormalization group (HOTRG)~\cite{hotrg2012}, many attempts to develop the 3D TNRG~\cite{atrg2020,triadtrg2019,Kadoh_2022} focus on reducing its computational complexity. 
However, it is not a higher computational complexity\footnote{
    The HOTRG has computational complexity $O(\chi^7)$ in 2D and $O(\chi^{11})$ in 3D.
} but the overflow of the microscopic entanglement that makes the 3D problem qualitatively different from the 2D one.
%
We must stress here that in 2D, the entanglement filtering was luxury for better accuracy, whereas it is basic necessity in 3D for any apt TNRG scheme\@.
Entanglement filtering in 3D should be manipulation of the tensor network to renormalize out the pre-factor $\alpha$ of the 3D area law in Eq.~\eqref{eq:area3dcft};
this indicates that the EDL structure in Fig.~\ref{fig:edl} provides a good guidance for the design of a better 3D TNRG.
In Ref.~\cite{Lyu:Kawashima:2024}, we propose a 3D entanglement-filtering scheme that can clear up the EDL structure, as well as a ``sphere-like'' structure~\cite{gilt2018}.
Numerical results on the 3D Ising model suggest that the scheme in Ref.~\cite{Lyu:Kawashima:2024} manages to tame the area-law term in Eq.~\eqref{eq:area3dcft}.

Finally, we want to point out some implications for entropic c-theorems~\cite{casini2007, nishioka2018}.
In quantum field theories, the relationship between RG and entanglement entropy takes the form of various entropic c-theorems.
The original c-theorem by Zamolodchikov~\cite{zomo-ctheorem} states that a central-charge-like $c$ quantity of 2D renormalizable quantum field theories is non-increasing along RG trajectories; 
$c$ is stationary at a fixed point and is proportional to the central charge of the corresponding conform field theory.
Entropic c-theorems generalize the central-charge-like quantity to the universal contribution to the entanglement entropy of a sphere for a general $d$-dimensional quantum field theory, taking the value $F$ in Eq.~\eqref{eq:area3dcft} in 3D.
In the context of an entropic c-theorem, the RG coarse graining is understood as increasing the radius of an entangling surface~\cite{nishioka2018}, in analogy to the block-tensor RG\@;
the divergent terms in the entanglement entropy similarly haunts the analysis~\cite{casini2015}.
One may hope that the idea of entanglement filtering developed in numerical TNRG and its recent rigorous realization~\cite{kennedy-hiT,kennedy-lowT} might inspire new ideas on how to deal with those divergent terms.

\backmatter

\bmhead{Acknowledgements}
We thank Slava Rychkov for clarifying the conformal-field-theory prediction of 3D Ising critical exponents.
We thank Raghav G. Jha and Zhiyuan Xie for explaining tensor-network RG analysis on 3D Ising and Potts model.
We thank Satoshi Morita for helping with the parallel computing implementation of 3D tensor-network RG.
We thank Shumpei Iino, Wei-Lin Tu, Katsuya Akamatsu, Kenji Homma, Antoine Tilloy, Takeo Kato, Hirokazu Tsunetsugu, Yuan Yao, Weiguang Cao and Shihwen Hor for fruitful discussions.
X.L.\ is grateful to the support of the Global Science Graduate Course (GSGC) program of the University of Tokyo.
The peer reviewing process of this work was done after X.L.\ moved to IHES.
This work is financially supported by JSPS KAKENHI Grant Number 23K25789.
The computation in this work has been done using the facilities of the Supercomputer Center, the Institute for Solid State Physics, the University of Tokyo.

\section*{Declarations}


\bmhead{Competing interests}
The authors have no competing interests to declare that are relevant to the content of this article.

\bmhead{Code availability}
The numerical results in this work can be reproduced using the codes published in the Ref.~\cite{Lyu:bkten3d:codes}.

\begin{appendices}
\section{Linearized RG map of the HOTRG}\label{secA1:linRGmap}
In the context of TNRG, the Wilsonian RG prescription for extracting scaling dimensions from the linearization of the RG map has been developed in Refs.~\cite{lyu-kawashima2021,Ebel:2024LDO}.
In this appendix, we demonstrate how to perform this prescription for the 2D HOTRG.
The generalization to the 3D HOTRG is straightforward and has been expounded in Ref.~\cite{lyu:kawashima:2025lattsym}.
The RG transformation of the 2D HOTRG~\cite{hotrg2012} is a nonlinear map $A \mapsto A'=\mathcal{T}(A)$, which is a composition of the following two collapses in two directions,
\begin{align}
    \label{eq:hotrg2step}
    \includegraphics[scale=1.0, valign=c]{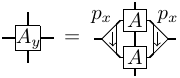},
    \includegraphics[scale=1.0, valign=c]{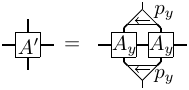}.
\end{align}
The order of the collapse is chosen arbitrarily to be $y \rightarrow x$ here.
The triangular tensors $p_x, p_y$ are known as isometric tensors and can be determined using either the higher-order singular value decomposition~\cite{hotrg2012}, or, equivalently, the projective truncations~\cite{Evenbly:tnralgo,lyu:kawashima:2025lattsym}.

In the first stage of the Wilsonian RG prescription, by tuning the temperature of the model, one determines a fixed-point tensor $A_*$ of the above RG map such that it is invariant under the RG map: $A_* = \mathcal{T}(A_*)$.
Due to the gauge freedom of a tensor-network representation of the partition function, a gauge-fixing scheme is necessary at this stage~\cite{lyu-kawashima2021,Ebel:2025Rot} for obtaining a fixed-point tensor $A_*$ in numerical calculations.
The extent to which the tensor $A_*$ is fixed numerically can be greatly improved by the Newton Method~\cite{Ebel:2025Rot}.

In the second stage, when $A_*$ has been determined, let us denote the corresponding isometric tensors to be $p_{x*}, p_{y*}$.
The linearized map of the HOTRG is a linear map $\delta A \mapsto \delta A' = \mathcal{R}(\delta A)$, which can be built from $A_*, p_{x*}, p_{y*}$ according to Refs.~\cite{lyu-kawashima2021,Ebel:2024LDO} as the following composition:
\begin{subequations}
    \label{eq:linhotrg}
\begin{align}
    \label{eq:linhotrg2step}
    \includegraphics[scale=0.9, valign=c]{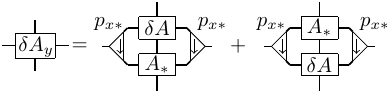},
    \includegraphics[scale=0.9, valign=c]{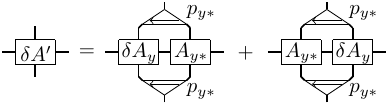},
\end{align}
where
\begin{align}
    \label{eq:AystarDef}
    \includegraphics[scale=1.0, valign=c]{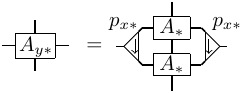}.
\end{align}
\end{subequations}
According to Wilsonian RG theory, the scaling dimensions $\{\Delta_i\}$ of the criticality can be extracted from the eigenvalues  $\{\lambda_i \}$ of the linearized RG map $\mathcal{R}$ in Eq.~\eqref{eq:linhotrg}:
\begin{align}
    \label{eq:linmap2scaleD}
    b^{d-\Delta_i} = \lambda_i,
\end{align}
where $b=2$ is the rescaling factor of the RG transformation and  $d=2$ is the spatial dimensionality of the system.

\end{appendices}


\bibliography{references}

\end{document}